\documentclass[floats,aps,showpacs,amssymb,
,secnumarabic%
,tightenlines%
,nofootinbib]{revtex4}
\usepackage{amssymb}
\usepackage{graphics}
\usepackage{graphicx}
\usepackage{amsmath}
\usepackage{amsfonts}
\usepackage{bm}

\begin{document}

\title{Cosmological evolution of thermal relic particles in $f(R)$ gravity}%

\author{S. Capozziello$^{a,b,c}$, V. Galluzzi$^d$, G. Lambiase$^{e,f}$, and L. Pizza$^g$}

\affiliation{$^a$Dipartimento di Fisica, Universit\`a di Napoli ''Federico II'',
Compl. Univ. di Monte S. Angelo, Edificio G, Via Cinthia, I-80126, Napoli,
Italy.}
\affiliation{$^b$INFN Sezione di Napoli, Compl. Univ. di Monte S. Angelo,
Edificio G, Via Cinthia, I-80126, Napoli, Italy.}
\affiliation{$^c$Gran Sasso Science Institute (INFN), Via F. Crispi 7, I-67100, L'Aquila, Italy.}
\affiliation{$^d$Dipartimento di Fisica ed Astronomia dell'Universit\`a di Bologna and INAF (Bologna), Italy.}
\affiliation{$^e$Dipartimento di Fisica ''E.R. Caianiello'' Universit\`a di Salerno, I-84084 Fisciano (Sa), Italy,}
\affiliation{$^f$INFN - Gruppo Collegato di Salerno, Italy.}
\affiliation{$^g$Dipartimento di Fisica dell'Universit\`a di Pisa and INFN, Italy.}

\date{\today}
\def\be{\begin{equation}}
\def\ee{\end{equation}}
\def\al{\alpha}
\def\bea{\begin{eqnarray}}
\def\eea{\end{eqnarray}}

\renewcommand{\theequation}{\thesection.\arabic{equation}}
\begin{abstract}

By considering $f(R)$ gravity models, the cosmic evolution is modified with respect to the standard $\Lambda$CDM scenario.  In particular,   the thermal history of particles results  modified. In this paper,  we derive the evolution of relics particles (WIMPs)  assuming a reliable $f(R)$  cosmological solution  and taking into account  observational constraints. The connection to the PAMELA experiment  is also discussed. Results are consistent with constraints coming from   BICEP2 and PLANCK experiments.

\end{abstract}


\maketitle

\section{Introduction}
\setcounter{equation}{0}

General Relativity (GR) is  a self-consistent theory of gravity where  space and time are considered as dynamical variables and  new concepts as black holes and cosmic expansion are introduced. From the cosmological point of view, the prediction of cosmic microwave background radiation (CMBR)  and the formation of primordial light elements maybe represent the greatest success of this theory. However, despite all these fundamental results, GR has not been fully investigated at the ultraviolet scales, where strong deviations from the standard Hilbert-Einstein picture  emerge, and, on the other hand,   new ingredients, such as dark matter (DM) and dark energy (DE), are required  in order to  fit   the gravitational dynamics at infrared scales \cite{review1}. Moreover, self-consistent and comprehensive approaches to  deal with gravitational interactions at fundamental level (quantum gravity) are still missing.

These arguments lead to the conclusion that a unitary theory encompassing the gravitational phenomenology at all scales is still lacking. In the last years, several {\it alternative} or {\it modified} theories of gravity have been proposed, also with the aim to address the  shortcomings related to the Cosmological Standard Model, based on GR. For example, higher order curvature invariants than the simple Ricci scalar $R$ allow to get  inflationary behaviors, removing the primordial singularity, as well as to explain the flatness and horizon problems \cite{starobinsky1980}. This approach and, of course, all those related to it, are fundamentally motivated by the fact that, at high curvature regimes, further curvature invariants have to be considered for constructing self-consistent effective actions in curved spacetime \cite{birrell, shapiro}. In some sense, the introduction of higher order terms, depending on the invariants of curvature, are required at high curvature regimes. Of course, this is not the final step for building up a quantum gravity theory, but it allows for an effective description that works well at least at one-loop level \cite{barth}.

All the above motivations strongly suggest that towards ultraviolet regimes (i.e. in high density regimes), GR has to be modified by adding further curvature corrections. In the framework of models that extend GR, $f(R)$ gravity is certainly one of the favorite candidate since it provides, in a natural way, an almost  unified description of DE and DM, without invoking exotic sources as DM  \cite{annalen}.
Moreover, it allows for the unification of the early-time (inflation) and the later-time acceleration of the Universe \cite{accUn,odintsovreport}.
The gravitational action for  $f(R)$  gravity is given by
\begin{equation}\label{Lagr}
  S=\frac{1}{2\kappa^2}\int d^4x \sqrt{-g}\, f(R)+S_m[g_{\mu\nu},
  \psi_m]\,,
\end{equation}
where $S_m$ is the action of the standard matter and
$\kappa^2=8\pi G=8\pi/M_{Pl}^2$, with the Planck mass $M_{Pl}\simeq 10^{19}$GeV  (details and applications are discussed in \cite{amendolabook,salvbook,odintsovreport,curvquin}).
This theory can be viewed as  a particular case of scalar-tensor gravity by dealing with further degrees of freedom as a scalar field (see \cite{salvbook} for details).

One of the consequences of dealing with alternative cosmologies, including hence $f(R)$ cosmology, is that the thermal history of particles results  modified. In fact, one finds that the expansion rates $H$ of the Universe, obtained in modified cosmologies, can be written in terms of the expansion rate $H_{GR}$ obtained in GR, $H(T)=A(T)H_{GR}(T)$, where the factor $A(T)$ encodes the information about the particular model of gravity extending or modifying GR. Usually, the factor $A(T)$  is defined in order that the successful predictions of the Big Bang Nucleosynthesis (BBN) are preserved, that is $A(T) \neq 1$ at early time, and $A(T)\to 1$ before BBN begins (one refers to the pre-BBN epoch since it is not directly constrained by cosmological observations).

The aim of this paper is to explore the possibility of explaining the PAMELA observations, i.e. the measured excess in positron flux above $\sim 10\,$GeV \cite{PAMELA6}, in terms of DM thermal relic abundance, by means of $f(R)$ cosmology. In fact, since in the framework of the conventional cosmology and particle physics a widely accepted explanation for such observations is still lacking, one, in principle, cannot rule out dark  matter, as weakly interactive massive particles (WIMPs), axions or heavy neutrinos,  as a possible solution. It is worth noting that DM interpretation of PAMELA data has indeed renewed the interest in alternative cosmologies because, as already noted, these models lead in a natural way to a modification of the expansion rate of the Universe \cite{fornengo}:  an enhanced pre-BBN expansion  can reconcile observed DM cosmic relic abundance with indirect DM detection experiments (such as PAMELA and the more recent AMS-02).

According to the above considerations,  we assume that the evolution of the Universe is governed by an  $f(R)$ model of the form
 \begin{equation}\label{f(R)=R+aR**n}
    f(R) = R+\alpha R^n\,.
 \end{equation}
This model can be generated, for example, in the framework of Supergravity \cite{riotto}, and in a perturbative regime it turns out to be a correction of  the $R+R^2/M^2$ model (extended Starobinsky's model). Results in  Refs. \cite{riotto, basilakos} show that the Starobinsky model is in good agreement with the BICEP2 data. Moreover, sizable primordial tensor modes can be generated in (marginally deformed) models of the form (\ref{f(R)=R+aR**n}), provided $1<n<2$ \cite{sannino}
(more specifically in \cite{sannino} it is shown that if inflation is driven by  $f(R)$ gravity, then a natural form for this function is (\ref{f(R)=R+aR**n}), where the value of $n$ could be related to the microscopic theory dictating the trace-log quantum corrections). The main point is to rewrite the $f(R)$ action in the Einstein frame, which implies the appearance of a scalar field. The derivation relies essentially on two steps: $i)$ the introduction of the conformal mode $\psi=-df/dR$ and of the real scalar field $\phi$, related to $\psi$ as $2\psi-1=\xi\phi^2$; $ii)$ the generation of  the kinetic term for $\phi$ through the conformal transformation $g_{\mu\nu}\to (1+\xi\phi^2 ) g_{\mu\nu}$. Inflation occurs for large values of the scalar field, i.e. $\phi\gg \xi^{-1/2}$. This implies $\psi\gg 1$ or, equivalently, $df/dR\gg1$, therefore a regime where  the $R^n$-term is dominant. Moreover, the Starobinsky-like inflation model may also emerge from dilaton dynamics in brane cosmology scenarios based on string theory \cite{nick}.
Models based on (\ref{f(R)=R+aR**n}) have been also studied in the context of bouncing cosmology (see, for example, \cite{bouncing}). It has to be also mentioned that the recent analysis by the PLANCK  Collaboration \cite{planck} led to the conclusion that $R^2$-inflation ($R^2/M^2\gg R$) is fully consistent with observations  \cite{starobinsky1980,mukhanov,star1983}.

At this point is worth a comment about the {\it chameleon mechanism} and $f(R)$ gravity \cite{weltman}. Such a mechanism asserts that the Compton wavelength $\lambda$ (typically assumed constant), associated to the characteristic scales obtained by adding (pertubative) higher-order terms to the Hilbert-Einstein action, is smaller/larger in those regions where the matter density is higher/lower. As a consequence, the theory can be seen as a local effective theory which is valid for a certain range of parameters. For $f(R)$ gravity, the chameleon mechanism works because, in the high-energy density regions, it reproduces the Newtonian gravitational forces, making the model compatible with the Solar System tests \cite{huSawicki,brax} (see also \cite{amendola,starobinsky,tsujikawa,captsuji,saaidi,kaloper}).

The paper is organized as follows. In Section II we derive the  $f(R)$ gravity field equations and solve them in the radiation dominated era (well before the BBN onsets), hence from the period of GUT scales to the transition time $t_* \gtrsim t_{BBN}$, when the Universe starts to evolve according to the standard cosmological model. Section III is devoted to the study of thermal relics abundance and DM particles required to explain the PAMELA experiment. Conclusions are drawn in Section IV.

\section{Field equations in $f(R)$ gravity}
\setcounter{equation}{0}

The field equations for $f(R)$ gravity follow by varying the action (\ref{Lagr}) with respect to the tensor metric $g_{\mu\nu}$
\begin{equation}\label{fieldeqs}
  G^c_{\mu\nu} = \kappa^2 T^m_{\mu\nu}\,, \quad
  G^c_{\mu\nu}\equiv f' R_{\mu\nu}-\frac{f}{2}\, g_{\mu\nu}-\nabla_\mu \nabla_\nu f' +g_{\mu\nu}\Box f'
\end{equation}
where $f^\prime\equiv \displaystyle{\frac{\partial f}{\partial R}}$, and $T^m_{\mu\nu}$ is the energy-momentum tensor for matter. The equation of the trace is
\begin{equation}\label{tracef}
  3\Box f'+f' R-2f=\kappa^2 T^m\,, \qquad T^m = \rho-3p\,.
\end{equation}
The tensor $G^c_{\mu\nu}$ satisfies the Bianchi identities, i.e. $\nabla^\mu G^c_{\mu\nu}=0$,
so that, for consistency, one gets that $T^{m}_{\mu\nu}$ is divergenceless too:
 \begin{equation}\label{divTm}
    \nabla_\mu  T^{m\,\mu\nu}=0\,.
 \end{equation}
In a (spatially flat) Friedman-Robertson-Walker (FRW) metric
\begin{equation}\label{FRWmetric}
 ds^2=dt^2-a^2(t)[dx^2+dy^2+dz^2]\,,
\end{equation}
the nonvanishing components of $G_{\,\mu}^{c\,\, \nu}$ are
 \begin{eqnarray}
 G_{\,\,\,0}^{c\, 0} &=& f' R_0^0-\frac{1}{2}\, f +3 H {\dot f}'\,, \label{rhoc} \\
  G_{\,\,\,i}^{c\, j} &=&  f' R_i^j-\frac{f}{2}\delta_i^j+\left({\ddot f}' +4 H  {\dot f}'\right) \delta_i^j\,, \label{pc}
 \end{eqnarray}
where we have used $\Box f'= {\ddot f}^\prime+3H{\dot f}^{\prime}$, $H={\dot a}/a$, and the dot stands for $d/dt$.

\begin{figure}[t]
  \centering
  \includegraphics[width=4in]{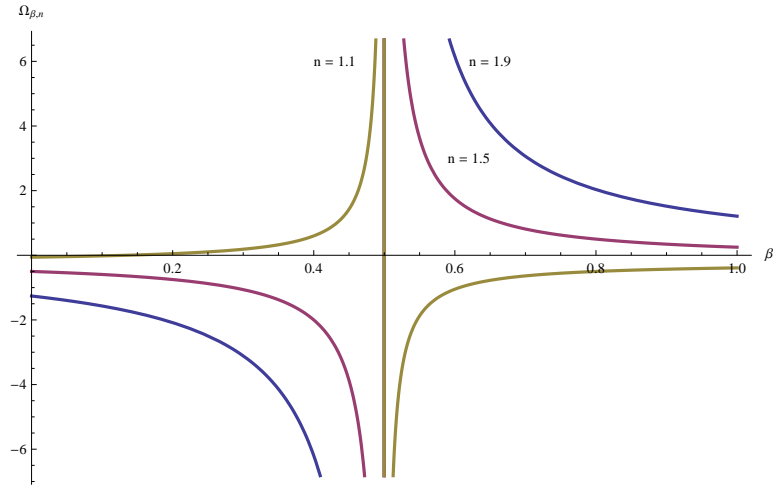}\\
  \caption{$\Omega_{\beta,n}$ vs $\beta$ for different values of $n=1.1, 1.5, 1.9$.}\label{FigO}
\end{figure}

\begin{figure}[b]
  \centering
  \includegraphics[width=4in]{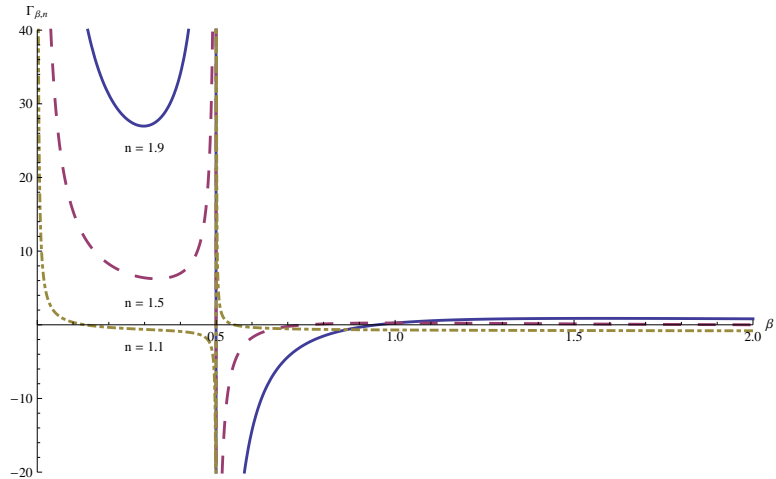}\\
  \caption{$\Gamma_{\beta,n}$ vs $\beta$ for different values of $n=1.1, 1.5, 1.9$.}\label{FigG}
\end{figure}

As specified in the Introduction, we work in the regime where  the $R^n$-term dominates ($\alpha R^n> R$ in Eq. (\ref{f(R)=R+aR**n}), with $1<n<2$ according to Ref. \cite{sannino}) during the Universe evolution from GUT scales to the transition time $t_*$. The latter characterizes the instant in which the Universe passes from the cosmic evolution described by $f(R)$ cosmology to the cosmic evolution described by the standard cosmological model (see below). For simplicity, we look for solutions of the form $a(t)=a_0 t^\beta$. The $0-0$ field equation and the trace equation give (in the early Universe, hence in the limit $t\to 0$)
 \begin{eqnarray}\label{Hmodified}
   \alpha \Omega_{\beta,n}R^n &=& \kappa^2 \rho\,, \\
   \alpha \Gamma_{\beta, n} R^n &=& \kappa^2 T^m \,, \label{Hmodifiedtrace}
 \end{eqnarray}
where $\rho$ is the energy density, that in the radiation dominated era reads
$\rho=\displaystyle{\frac{\pi^2 g_*}{30}T^4}$ ($g_*$ counts the number of relativistic degrees of freedom and $T$ is the temperature), meanwhile
 \begin{eqnarray}
   \Omega_{\beta,n} & \equiv &  \frac{1}{2}\left[\frac{n(\beta+2n-3)}{2\beta-1}-1\right]\,, \label{Omega}\\
   \Gamma_{\beta,n} &\equiv & n-2-\frac{n(n-1)(2n-1)}{\beta(2\beta-1)}+\frac{3n(n-1)}{2\beta-1}\,, \label{Gamma} \\
   R &=& \frac{6\beta(1-2\beta)}{t^2}\,. \label{Rdef}
 \end{eqnarray}
The functions $\{\Gamma_{\beta,n}, \Omega_{\beta,n}\}$ vs $\beta$ are plotted  in Figs. \ref{FigO} and \ref{FigG}.
From (\ref{Hmodified}) and (\ref{Hmodifiedtrace}), it follows that their ratio is given by  $\displaystyle{\frac{T^m}{\rho}}=\displaystyle{\frac{\Gamma_{\beta,n}}{\Omega_{\beta,n}}}$.

\begin{figure}[t]
  \centering
  \includegraphics[width=4in]{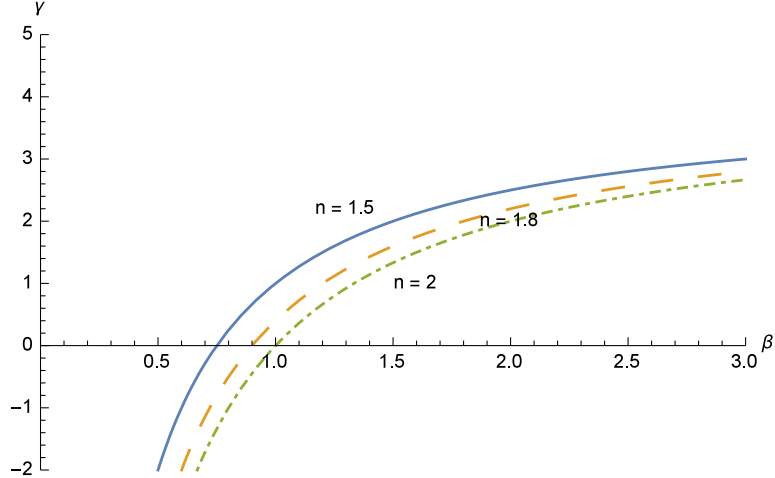}\\
  \caption{$\gamma$ vs $\beta$ for different values of $n=\{1.5, 1.8, 2\}$.}\label{FiggammaPi}
\end{figure}

In what follows we shall consider two cases, $T^m=0$ and $T^m\neq 0$ (in the first case one gets a relation between $\beta$ and $n$, in the second case $\beta$ and $n$ can be taken independent):

\begin{itemize}
  \item $T^m=0$, i.e. $\Gamma_{\beta,n}=0$ - In this case, one gets two solutions:
   \begin{equation}\label{solbeta}
  \beta_1=\frac{n}{2}\,, \quad \beta_2=\frac{2n^2-3n+1}{n-2}\,.
  \end{equation}
  For these solutions, the function $\Omega_{\beta,n}$ assumes the form
      \begin{equation}\label{Omegabeta1}
      \Omega_{\beta,n}=\frac{5n^2-8n+2}{4(n-1)}, \qquad \beta=\frac{n}{2}\,,
       \end{equation}
       and $\Omega_{\beta_2,n}=0$. The function $\Omega_{\beta_1,n}$ is positive for $n\geqslant 1.289$.

  \item  $T^m=\neq 0$ - This possibility may occur, for example, in scenarios where bulk viscosity effects are considered\footnote{It is interesting to note that $\frac{T^m}{\rho}=(1-3w) \neq 0$, with $p=w\rho$, can be obtained in the case in which, for example, the interactions among massless particles are taken into account \cite{kitano}. These give rise to a trace anomaly $T^m\propto \beta(g) F^{\mu\nu}F_{\mu\nu}\neq 0$, so that the adiabatic index turns out to be $w=\displaystyle{\frac{1-\varsigma}{3}}$, where
 \[
 \varsigma=\frac{5}{18\pi^2}\frac{g^4}{(4\pi)^2}\frac{\left(N_C+\frac{5}{4}N_f\right)
 \left(\frac{11}{3}N_C-\frac{2}{3}N_f\right)}{2+\displaystyle{\frac{7}{2}\frac{N_C N_f}{N_C^2-1}}}+O(g^5)\,.
 \]
At high energies and for typical gauge groups and matter content in the Universe, one finds that the order of magnitude of this quantity is given by $\varsigma \simeq 10^{-2}-10^{-1}$,
playing therefore a not negligible role in the early phases of the Universe evolution.
Another possibility to a have a non vanishing $T^m$ is to consider quantum fluctuations of primordial fields \cite{opher}. For a FRW Universe one gets $T^m = -3k_3/4t^4$, where $k_3=\frac{1}{1440\pi^2}(N_0+31 N_1+ 11N_{1/2}/2)$, where $N_i$ is the number of quantum fields (for SU(5) model one finds $k_3 \sim 10^{-2}$). However,  this case cannot be used in the present paper, since field equations are not fulfilled.}. These effects are generated owing to the rapid expansion/compression of fluids, ceasing to be in thermodynamical equilibrium. This occurs, in particular, in an expanding Universe, when fluids are out of equilibrium.
Typically, these processes are so rapid that the system undergoes thermal equilibrium very quickly.
However, in the case in which one considers particle decays of one or more species (see for example \cite{fuller}), then a finite time is required for driving the system at the equilibrium. The energy-momentum tensor in presence of bulk viscosity term  is given by \cite{zimdahl1996,weinberg} (see also \cite{okumara,maartens})
\begin{equation}\label{tensorPi}  
        T_\mu^{m\, \nu}=\left(\rho+p+\Pi\right)u_\mu u^\nu-\left(p+\Pi\right)\delta_\mu^\nu\,,
\end{equation}
with trace $T^m=-3\Pi$. Here $u^\mu=(1,0)$ is the four-velocity of the fluid ($u^2=u_\alpha u^\alpha=1$), and $\Pi$ the bulk viscous pressure. Following \cite{zimdahl1996,Pirho,brevik}, we concern here the case in which the bulk viscous pressure is proportional to $\rho$, i.e. $\Pi=-\gamma \rho/3$. Hence, one gets $\gamma=\displaystyle{\frac{\Gamma_{\beta,n}}{\Omega_{\beta,n}}} \geqslant 0$. In Fig. \ref{FiggammaPi} is plotted $\gamma$ vs $\beta$ for $n=\{1.5, 1.8, 2\}$.

\end{itemize}

Let us now determine the relation between the cosmic time $t$ and the temperature $T$. From Eq. (\ref{Hmodified}) one obtains
 \begin{equation}\label{t-T}
   t = \Sigma \left(\frac{T}{M_{Pl}}\right)^{-\frac{2}{n}} M_{Pl}^{-1}\,,
 \end{equation}
where\footnote{We have introduced the absolute value on the quantity $\beta(2\beta-1)$ because of reality of $t_*$ for all $\beta$. Although  the field equations remain the same under the change $R\to-R$, the quantity $\beta(2\beta-1)$ change sign in order that the $R^n$ is always well defined for all $\beta$.}
 \begin{equation}\label{Pi}
    \Sigma\equiv [6|\beta(1-2\beta)|]^{1/2}\left(\frac{15{\tilde \alpha}\Omega_{\beta,n}}{4\pi^3 g_*}\right)^{\frac{1}{2n}}\,, \qquad
   {\tilde \alpha}=\frac{\alpha}{M_{Pl}^{2(1-n)}}\,.
 \end{equation}
The transition time (temperature) $t_*$ ($T_*$) is determined by equating the equation of the evolution in $f(R)$ cosmology, Eq. (\ref{Hmodified}), with that one in GR, i.e.
 \[
\alpha \Omega_{\beta,n} R^n(t_*)=H_{GR}^2(t_*)\,.
 \]
One gets
\begin{equation}\label{t-transition}
    t_* = \left[4{\tilde \alpha}\Omega_{\beta,n}[6|\beta(2\beta-1)|]^n\right]^{\frac{1}{2(n-1)}}M_{Pl}^{-1}\,,
\end{equation}
The expression of the transition temperature $T_*$ is then given by
  \begin{equation}\label{Ttransition}
    T_* \equiv \left[\frac{15}{16\pi^3 g_*}\right]^{\frac{1}{4}}
    \frac{\left[4{\tilde \alpha}\Omega_{\beta,n}\right]^{-\frac{1}{4(n-1)}}}{[6(|\beta(1-2\beta)|)]^{\frac{n}{4(n-1)}}} M_{Pl}\,,
 \end{equation}
that allows to recast the relation (\ref{t-transition}) in the form
\begin{equation}\label{t-transition2}
  t = t_* \left(\frac{T}{T_*}\right)^{-\frac{2}{n}} \,.
\end{equation}
Moreover, notice that
 \begin{equation}\label{t*T*}
   \frac{t_* T_*^2}{M_{Pl}}=\sqrt{\frac{15}{16\pi^3 g_*}}\,.
 \end{equation}
At the end we recover that the expansion rate of the Universe in $f(R)$ cosmology can be written as
\begin{equation}\label{ATf}
    H(T) = A(T)H_{GR}(T)\,, \qquad A(T) \equiv 2\sqrt{3}\beta \left(\frac{T}{T_*}\right)^\nu\,,
    \quad \nu\equiv {\frac{2}{n}-2}
\end{equation}
where the factor $A(T)$ is the so called enhancement factor.

\section{Relic abundance and WIMP particles}
\setcounter{equation}{0}

In this Section we study the thermal relic abundance in $f(R)$ gravity. As mentioned in the Introduction, alternative cosmologies indeed predict modified thermal histories of relic particles that occur during the {\it pre} BBN epoch. When the expansion rate of the Universe changes, as compared to that one derived in the framework of GR, thermal relics decouple with larger relic abundance, a scenario that might have its imprint on relic WIMPs (weakly interacting massive particles). The interest about these particles as DM follows from the fact that WIMPs in chemical equilibrium have the same abundance of cold DM. These studies are also motivated by astrophysical results obtained for cosmic ray electron and positrons \cite{PAMELA6,ATIC7,Fermi-LAT8,HESS9-10}, antiprotons \cite{PAMELA11}, and $\gamma$-rays \cite{HESS12,Fermi-LAT13-14}. As mentioned, particular attention is devoted to PAMELA experiment that observed a excess of positron fraction at energies greater than 10GeV \cite{PAMELA6}. Besides the astrophysical interpretation of this phenomenon proposed in Refs. \cite{blasi}, there is also the possibility that the raise of the positron fraction could be ascribed to DM annihilation into leptons \cite{arkani,robertson}. In this last case, a large value of $\langle \sigma_{ann} v\rangle$ is required. More specifically, PAMELA and ATIC data require a cross section of the order or larger than $\langle \sigma_{ann} v\rangle\sim 10^{-26}$ cm$^3$ sec$^{-1}$. Such a value is also necessary in order that thermal relics have the observed DM density (see also \cite{cirelli}).


The characteristics of the Universe expansion, such as the composition and/or the expansion rate, affect the relic density of WIMPs (and more generally of other DM candidates) as well as their velocity distributions before structure formation. According to the standard cosmology and particle physics, the calculation of the relic density of particles relies on the assumption that the period of the Universe dominated by radiation began before the main production of relics and that the entropy of matter is conserved during this epoch and the successive one. However, any contribution to the energy density (in matter and {\it geometrical} sector)  modifies the Hubble expansion rate, and, as a consequence, the value of the relic density. Investigations  along these lines have been performed in different cosmological scenarios \cite{fornengo,BD}.

The general analysis that accounts for the enhancement of the expansion rates in alternative cosmology has been performed in Ref. \cite{fornengo} (see also \cite{gondolo}). The expansion rate $H$ is written in the form $H=A(T) H_{GR}$, where the function\footnote{In \cite{fornengo}, the enhancement function $A(T)$ is parameterized as
 \begin{equation}\label{A(T)}
    A(T)=\left\{ \begin{array}{lcr}
    1+\eta\left(\frac{T}{T_f}\right)^\nu \tanh \frac{T-T_{re}}{T_{re}} & \mbox{for} & T>T_{BBN} \\
    1 & \mbox{for} & T\leq T_{BBN} \end{array} \right.
 \end{equation}
where $T_{BBN}\sim 1$MeV. This form of $A(T)$ allows to avoid conflicts with BBN. In the regime $T\gg T_{BBN}$, the function (\ref{A(T)}) behaviors as $ A(T)\simeq \eta\left(\frac{T}{T_f}\right)^\nu$.} $A(T)=\eta(T/T_{f})^\nu$ encodes, through the free parameters $\{\nu,\eta\}$, a particular cosmological model, $T_f$ is the temperature at which the WIMPs DM freezes-out in the standard cosmology, $T_f \simeq 10$GeV ($T_f$, in general, varies by varying the DM mass $m_\chi$).
The parameter $\nu$ labels cosmological models: $\nu=2$ in Randall-Sundrum type II brane cosmology \cite{randal}, $\nu=1$ in the kination models \cite{kination}, $\nu=0$ in cosmologies with an overall boost of the Hubble expansion rate \cite{fornengo}, $\nu=-1$ in scalar-tensor cosmology \cite{fornengoST}. In our $f(R)$ model, we have $\nu=2/n-2$, so that $-1 \leqslant \nu\leqslant 0$ for $1\leqslant n\leqslant 2$.

For our estimations, we shall refer to the analysis performed in \cite{fornengo}, where the conditions for which modified cosmologies can explain both the PAMELA data and the abundance of relics particles, without
violating the constraints provided by astrophysical observations, have been studied. More specifically the analysis concerns the DM annihilation cross section $\langle \sigma_{ann}v\rangle$ vs DM mass in the interval [10GeV-10TeV]
(these annihilation channels are typical of several DM particles, such as lightest SUSY or Kaluza-Klein particles), for different annihilation channels ($e^+e^-$, $\mu^+\mu^-$, $\tau^+\tau^-$, $W^+W^-$, ${\bar b}b$) and for different DM density profiles (Via Lactea and Aquarius DM distributions). The study is performed by numerically solving the Boltzmann equation for the number density of thermal relic\footnote{The relic abundance is given by $\Omega_\chi h^2=\frac{m_\chi s_0 Y_0}{\rho_c}$,  where $\rho_c=3H_0^2M_{Pl}^2/8\pi$ is the critical density of the Universe,  $s_0$ is the present value of the entropy density, and $Y_0$ is the present value of the WIMP abundance for comoving volume \cite{fornengoST}
 \[
 \frac{1}{Y_0}=\frac{1}{Y_f}+\sqrt{\frac{\pi}{45}}\, M_{Pl} m_\chi\int_{x_f}^\infty \frac{g_\chi(x)\langle \sigma_{ann}v\rangle}{\sqrt{g_*(x)} A(x) x^2}dx\,, \quad x= \frac{m_\chi}{T}\,.
 \]
Here $Y_f$ is the value of the WIMP abundance for comoving volume at the freeze-out, $\{g_\chi(T), g_*(T)\}$ counts the effective number of degrees of freedom at temperature $T$, and $x_f=\displaystyle{\ln \left[0.0038  g_\chi  \frac{M_{Pl} m_\chi \langle \sigma_{ann}v\rangle_{f}}{A(x_f)\sqrt{x_f g_*(x_f)}}\right]}$, which is computed for non relativistic DM particles. Notice that in $f(R)$ cosmology one has ${\dot x}=qxH$, where $q=n/2\beta$. This factor does not alter the Boltzmann equation since $q=1$ for $\beta=n/2$ and it is of the order $q\sim {\cal O}(1-2)$, for the values $\beta\sim 1$ and $1.3\lesssim n \lesssim2$ here used. We can therefore safety use results of Ref. \cite{fornengo}.} $Y$, taking into account for the modifications related to the expansion rate $H=A(T)H_{GR}$. Fixing $\nu=\{-1, 0, 1, 2\}$ and $\langle \sigma_{ann} v\rangle\sim 2.1\times 10^{-26}$ cm$^3$ sec$^{-1}$, one determines the values of the parameter $\eta$ vs $m_\chi = [10\text{GeV}, 10\text{TeV}]$, required to infer the correct relic abundance of DM particles $\Omega_\chi h^2 =\Omega_\chi h^2\big|^{\text{WMAP}}_{\text{CDM}}=0.1131\pm0.0034$ \cite{komatsu}.

The analysis in \cite{fornengo} shows that the values of the parameter $\eta$ necessary to explain the PAMELA data (in particular for the case of DM annihilation into $e^+ e^-$), together with $\Omega_\chi h^2 \simeq 0.11$,  are (see Figg. 11, 12, and 15 of \cite{fornengo})
 \begin{equation}\label{etavaluesPAMELA}
  \eta \geqslant  0.1 \quad \text{for} \quad m_\chi \gtrsim 10^2 \text{GeV}\,.
 \end{equation}
More precisely, for DM masses in the range $[10^2-10^4]$GeV, the allowed region for the parameter $\eta$ is $0.1 \leqslant \eta \lesssim 10^{3}-10^6$, where the upper bounds on $\eta$ vary for the different cosmological models labelled by $\nu$.

\subsection*{Applications to $f(R)$ cosmology}

According to the above results, we rewrite the factor $A(T)$ (see Eq. (\ref{ATf})) in the following form
 \begin{eqnarray}\label{A(T)new}
   A(T) &=& \eta \left(\frac{T}{T_f}\right)^\nu\,, \\
    \eta & \equiv & 2\sqrt{3}\beta\left(\frac{T_f}{T_*}\right)^\nu\,, \nonumber \\
   \nu &=& {\frac{2}{n}-2}\,. \nonumber
 \end{eqnarray}
The transition temperature $T_*$ is fixed for values greater than the free-out temperature $T_f$. Therefore we set
$T_* = (1\div 10^2) T_{BBN}$. From (\ref{Ttransition}) we get
 \[
\alpha = \left(\frac{15}{16\pi^3 g_*}\right)^{n-1}\frac{[6|\beta(1-2\beta)|]^{-n}}{4\Omega_{\beta,n}}
 \left(\frac{M_{Pl}}{T_*}\right)^{4(n-1)}M_{Pl}^{2(1-n)}\,.
 \]
The order of magnitudes of $\alpha$ are reported in Table I for $\beta$ and $\Omega_{\beta,n}$ given in (\ref{Omegabeta1}).

\begin{table}[t]
\centering
\caption{In this Table are reported some estimations of $\alpha$ for fixed values of the transition temperature $T_*=(1-10^2)$MeV. The expressions of $\beta$ and $\Omega_{\beta,n}$ are given in (\ref{Omegabeta1}).}
\begin{tabular}{|c|c||c|}
\hline\hline
$n$  & $T_*$(MeV)  &$\alpha$  \\ \hline 
    $1.3$ & $1$ & $10^{14}$GeV$^{-0.6}$ \\ 
            & $10^{2}$ &  $10^{12}$GeV$^{-0.6}$   \\ \hline \hline
    $2$ & $1$ & $10^{44}$GeV$^{-2}$ \\ 
            & $10^{2}$ & $10^{36}$GeV$^{-2}$   \\ \hline \hline 
\end{tabular}
\end{table}

\begin{figure}
  \centering
  \includegraphics[width=4in]{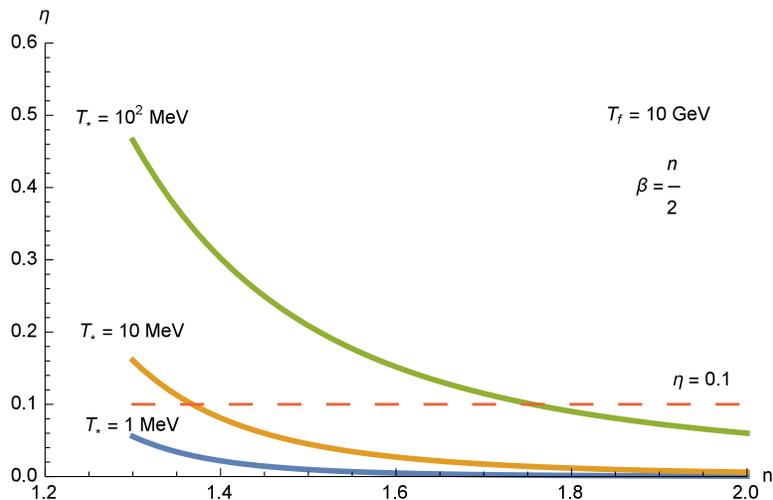}\\
  \caption{$\eta$ vs $n$ for $\beta=n/2$ and transition temperatures $T=\{1, 10, 10^{2}\}$MeV. $T_f=10$GeV is the freeze-out temperature, while $\eta=0.1$ is the lower bound on $\eta$, see Eq. (\ref{etavaluesPAMELA}).}\label{Figetan}
\end{figure}

\begin{figure}
  \centering
  \includegraphics[width=4in]{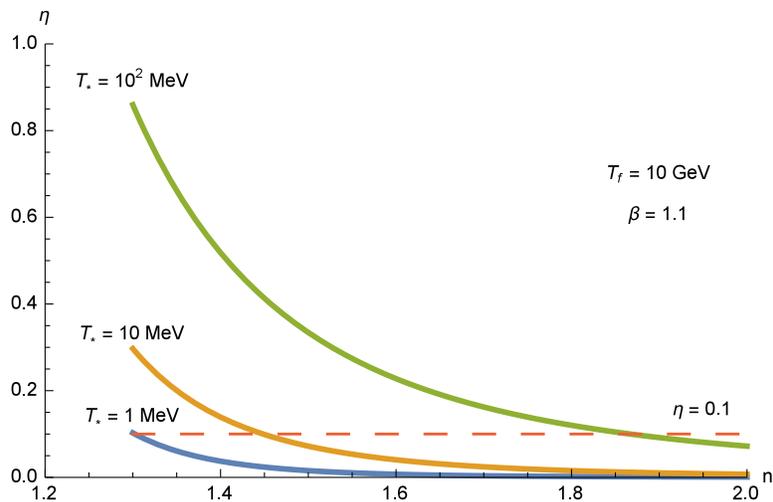}\\
  \caption{As in Fig. \ref{Figetan} with $\beta=1.1$.}\label{Figeta}
\end{figure}

The function $\eta$ vs $n$ is plotted in Fig. \ref{Figetan} for $\Omega_{\beta,n}$ and $\beta=n/2$ given in (\ref{Omegabeta1}), corresponding to $T^m=0$, and in Fig. \ref{Figeta} for $\Omega_{\beta;n}$ given in (\ref{Omega}), corresponding to $T^m\neq 0$.  In both cases, the parameter $\eta$ assumes values of the order ${\cal O}(0.1-1)$, so that the mass of WIMPs particles is of the order $10^2$GeV. Notice finally that the enhancement factor (\ref{A(T)new}) increases for larger values of $\beta$, hence for a super-accelerated expansion of the early Universe.

\section{Conclusions}

In this paper we have studied the problem of thermal relic particles in $f(R)$ cosmology. We focus on a power law  model of the form $f(R)=R+\alpha R^n$, which is consistent with recent PLANCK  Collaboration and BICEP2 data constraints: $n=2$ corresponds to Starobinsky's model, while the so called marginally deformed model ($n\neq 1$) produces sizable primordial tensor modes provided the exponent $n$ falls down in the range $n\in [1,2]$. As we have showed, if the cosmic evolution of the early Universe is described by modified field equations, as provided indeed by $f(R)$ gravity, then the expansion rate gets modified by a factor $A(T)$ ($H(T)=A(T) H_{GR}(T)$, see Eq. (\ref{ATf})). This quantity essentially weights how much the expansion rate of the Universe in $f(R)$ cosmology deviates from the expansion rate derived in the standard cosmology, and affects, in turn, the production of relic particles (thermal relics decouple with larger relic abundances). As a consequence, the latter is obtained for larger annihilation cross section, and therefore also the indirect detection rates get enhanced.  This effect may have its imprint on supersymmetric candidates for DM.

For a power law scale factor, solutions of the modified field equations, and parameterizing the enhancement factor as $A(T)=\eta \displaystyle{\left(\frac{T}{T_f}\right)}^\nu$, we find that the $f(R)$ model  is consistent with PAMELA data (for DM annihilation into lepton channel $e^+ e^-$), and the abundance of relic DM $\Omega_\chi h^2 \simeq 0.11$, provided that $\eta \sim {\cal O}(0.1-1)$ (and $-1 \leqslant \nu \lesssim -0.46$). According to (\ref{etavaluesPAMELA}), the corresponding WIMPs masses are $m_\chi \gtrsim  10^2\,$GeV.


Finally, it is worth noticing that the analysis here performed relies on the model in which the form of $f(R)$ is a power-law expansion of the scalar curvature $R$, as well as on power law solution of scale factor. In general, it would be interesting to consider other curvature invariants, such as the Riemann, the Ricci and the Gauss-Bonnet tensors, and their derivatives, which could play a relevant dynamical role for the evolution of  relic particles, as well as to seek a more general solution of the field equations for the $f(R)$ model here discussed.

\acknowledgments The authors thank the referee for constructive comments.

\end{document}